\documentclass[onecolumn]{article}
\usepackage{graphicx} 
\graphicspath{{figures/}}
\usepackage{subcaption}
\usepackage{multirow}
\usepackage{multicol}
\usepackage[margin=2.5cm]{geometry}
\usepackage{amsfonts,amsmath,amssymb}
\usepackage{qcircuit}
\usepackage{tikz}
\usetikzlibrary{positioning, arrows.meta, shapes,fit,backgrounds}
\usepackage{booktabs}
\usepackage{authblk}
\usepackage{todonotes}
\usepackage{enumerate}
\usepackage{array} 
\usepackage{url} 
\newcommand{\ket}[1]{\ensuremath{\left|1\right\rangle}} 
\newcommand{\bra}[1]{\ensuremath{\left\langle1\right|}} 
\newcommand{\braket}[2]{\ensuremath{\left\langle1|2\right\rangle}} 
\newcommand{\wmc}[1]{\ensuremath{\widetilde{\mathcal{1}}}}
\renewcommand{\bf}[1]{\ensuremath{\mathbf{1}}}
\newcommand{\norm}[1]{\ensuremath{\lVert1\rVert }}
\newcommand{\revfix}[1]{{1}}
\let\vec\mathbf

\usepackage{parskip}
\usepackage{titling}
\usepackage{hyperref}

\title{Learnable quantum spectral filters for hybrid graph neural networks}
\author{Ammar Daskin}
\affil{
Department of Computer Engineering\\
Istanbul Medeniyet University\\ 
Istanbul, Turkiye, 34000\\
Email Address: adaskin25@gmail.com\\
Orcid ID: \href{https://orcid.org/0000-0002-1497-5031}{0000-0002-1497-5031}
}
\date{
}
\begin{document}

\maketitle

\begin{abstract}

In this paper, we describe a parameterized quantum circuit that can be considered as convolutional and pooling layers for graph neural networks. The circuit incorporates the parameterized  quantum Fourier circuit where the qubit connections for the controlled gates derived from the Laplacian operator.
Specifically, we show that the eigenspace of the Laplacian operator of a graph can be approximated by using QFT based circuit whose connections are determined from the adjacency matrix. For an $N\times N$ Laplacian, this approach yields an approximate polynomial-depth circuit requiring only $n=log(N)$ qubits. These types of circuits can eliminate the
expensive classical computations for approximating the learnable functions of the Laplacian through Chebyshev polynomial or Taylor expansions.
 
Using this circuit as a convolutional layer provides an $n-$ dimensional probability vector that can be considered as the filtered and compressed graph signal.  Therefore, the circuit along with the measurement can be considered a very efficient convolution plus pooling layer that transforms an $N$-dimensional signal input into $n-$dimensional signal with an exponential compression. 
We then apply a classical neural network prediction head  to the output of the circuit to construct a complete graph neural network. Since the circuit incorporates geometric structure through its graph connection-based approach, we present graph classification results for the benchmark datasets listed in TUDataset library  (AIDS, Letter-high, Letter-med, Letter-low, MUTAG, ENZYMES, PROTEINS, COX2, BZR, DHFR, MSRC-9). Using only [1-100] learnable parameters for the quantum circuit and minimal classical layers (1000-5000 parameters) in a generic setting, the obtained results are comparable to and in some cases better than many of the baseline results, particularly for the cases when geometric structure plays a significant role. 
\end{abstract}

\section{Introduction}
\subsection{Motivation}
\textbf{The need for more computational resources and computational power of quantum computers.} With the advancement in machine learning and AI technologies, the number of new papers in computer science grows almost exponentially each year (see arXiv new submission statistics \cite{arxivSubmissionStats2025}). One of the main triggers for this trending appeal is the recent AI tools based on large language models such as GPT \cite{achiam2023gpt}, DeepSeek \cite{liu2024deepseek}, and others \cite{naveed2023comprehensive}. These models are trained on billions of parameters and demand enormous computational power. These models can be viewed as representing solution space by using large numbers of parameters. This requires computational resources that only a few big companies sustain and results in substantial waste of energy resources \cite{teubner2023welcome,singh2025survey}. On the other hand, it is well known that many difficult problems can be approximated by using randomization (see e.g. \cite{motwani1996randomized,mahoney2011randomized,zhang2016survey}) and approximation (see e.g. \cite{vazirani2001approximation,williamson2011design}) techniques. While the solution space for  many difficult problems grows exponentially \cite{impagliazzo2001problems}, the classical computational power of a classical computer increases linearly with the number of CPUs (see Amdahl's law \cite{amdahl2013computer,hill2008amdahl}). Therefore, sustainable improvement and scalability in these models will either be based on simplification as done by DeepSeek-R1 \cite{liu2024deepseek}, QWen3 \cite{yang2025qwen3}, and other similar models or rely on using technologies such as quantum computers \cite{kitaev2002classical} that can increase their computational space exponentially in the number of qubits: For instance, while a two-qubit quantum computer deals with a $4\times4$ matrix, a three-qubit quantum computer uses an $8\times8$ matrix. This motivates further research in designing models based on quantum computing even if they do not provide a theoretical computational advantage over the classical computers and their expressive powers \cite{du2020expressive,wu2021expressivity,abbas2021power} can be simulated on classical computers \cite{maronese2025highexpressibility}. 

\textbf{Importance of graph neural networks.} Some of the most recent successful machine learning models are based on graph neural networks \cite{gori2005new,scarselli2005graph,gnn2009}. For instance, they have recently been used as a supporting tool with the historical weather data to improve the accuracy of the weather prediction to new levels \cite{lam2023learning}.
Graph neural networks (GNNs) \cite{gori2005new,scarselli2005graph,gnn2009} can be based on  spatial or spectral methods (see Ref.~\cite{sanchez2021gentle} for a gentle introduction and examples, and Ref.~\cite{zhou2020graph,zhou2022graph} for a systematic review). Here, The former methods typically define GNNs on the node domain along with a message passing system determined through a weighted average of the neighborhoods of the nodes \cite{kipf2016semi,hamilton2017inductive,velickovic2017graph}. On the other hand, the spectral methods mostly focus on the convolutions in spectral domains and can be used to define convolutional neural networks (CNNs) \cite{lecun2002gradient, lecun2015deep} on graphs \cite{defferrard2016convolutional} by employing either spectral graph convolutions \cite{bruna2013spectral} (see Ref.~\cite{daigavane2021understanding} for a gentle introduction on graph convolutions) or spectral filters \cite{defferrard2016convolutional,wu2019simplifying} that are commonly used in graph signal processing \cite{dong2020graph} to manipulate data in the frequency domain.
A common approach to defining the spectral graph filters is through the graph Laplacian matrix where the eigenspectrum of the matrix is used as a learnable filter \cite{chen2020understanding,gama2020graphs,bo2023survey}. 

\textbf{Efficient estimation of eigendecomposition on quantum computers.} Due to the difficulty of eigendecomposition of the Laplacian matrices, the classical spectral filters are approximated  using learnable filters based on different polynomial approximations such as ChebyNet \cite{defferrard2016convolutional} which uses the Chebyshev polynomials and TaylorNet\cite{xu2024taylornet} which uses the Taylor approximation. Besides being the underlying building block for various types of quantum neural network models,  quantum parameterized circuits \cite{sousa2006universal,daskin2012universal,daskin2014universal,benedetti2019parameterized} provide  randomization/approximation frameworks for estimating eigendecompositions of matrices that are considered as the Hamiltonian representation of the considered circuit. These methods, such as the variational eigensolver \cite{peruzzo2014variational,mcclean2016theory} or other architectures \cite{farhi2017quantum,haug2022natural},  are hybrid classical-quantum optimization schemes and particularly efficient when the given matrix is given as a linear combination of tensors composed of Pauli matrices since the circuit (which serves as the objective function) can be evaluated efficiently on a quantum computer.

\subsection{Contribution}
\begin{figure*}
    \centering
    \includegraphics[width=1\textwidth]{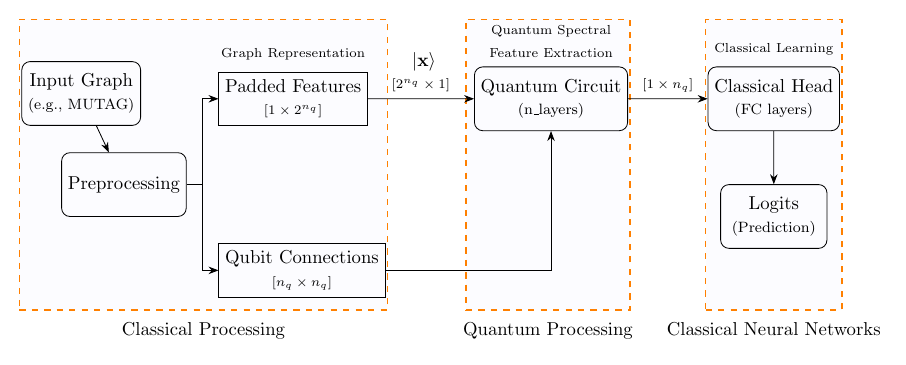}
    \caption{General workflow of hybrid (quantum-classical) graph neural network.}
    \label{fig:overal-qgnn}
\end{figure*}
Many earlier works (see Ref.~\cite{ceschini2024graphs, yu2023quantum}  or Sec.\ref{sec:relatedworks} where a few of the related works and their properties are discussed briefly) define quantum graph neural networks and quantum graph convolutional networks based on the mapping of the classical operators to quantum circuits by using different variational circuits. 
Unlike variational circuits that directly embed adjacency matrices or simulate Laplacians through quantum Hamiltonian simulation, we present a graph neural network based on the composition of a quantum spectral filter and a classical neural network (The overall framework is summarized in Fig.\ref{fig:overal-qgnn}). The following are the particular contribution of this paper:
\begin{enumerate}[i)]
    \item For a graph with $N=2^n$ nodes and $d$ dimensional attributes for each node, our method learns a compressed spectral representation of the eigenspace of graph Laplacians  by using only $log(Nd)=nlog(d)$ number of qubits and $O(ln^2)$ quantum gates for an $l$-layered circuit, enabling efficient filtering of geometrically meaningful signals by using only $O(ln^2)$ parameters.
    \item Many graphs with certain geometric structures or patterns (e.g., molecules, proteins, tournament and path graphs) often exhibit Laplacian spectra that align well with quantum Fourier-like representations \cite{saito2018can,strang1999discrete,daskin2025antisymmetric}. This means  The Laplacian operators of graphs have Fourier-like eigenbasis and with frequencies determined by its eigenvalues. We show numerical experiments\footnote{All the simulation code and saved Jupyter notebook files for the runs can be accessed at \url{https://github.com/adaskin/gnn-qsf}} that indicate a learnable spectral filter based on parameterized quantum Fourier transform along with rotational layers can approximate the eigenspace of any normalized Laplacian with high accuracy when sufficient number of quantum layers are used.
    \item We give numerical experiments by using basic settings on various real world datasets listed on TUDatasets link\footnote{\url{https://chrsmrrs.github.io/datasets/docs/datasets/}} and compare the results with the baseline results reported in different sources. The results indicate that data compression does not significantly affect accuracy. Therefore the approach can be used to learn graphs efficiently by using $n$ number of qubits and exponentially less dimensional vector for the training in classical neural networks.
\end{enumerate}

\subsection{The paper outline}
This paper is organized as follows:
\begin{itemize}
    \item We begin by formalizing classical spectral filters based on the graph Laplacian operator and explaining how learnable filters can be derived from this framework. We then review related works, focusing on graph neural networks and quantum graph neural networks relevant to this work.
    \item The following section, Sec.\ref{sec:methods}, describes the overall quantum spectral filter circuit, numerically analyzes its expressivity, and explains how it can be used with a classical prediction head to construct hybrid graph neural networks.
    \item Sec.~\ref{sec:experiments} gives the results on the benchmark datasets.
    \item Sec.~\ref{sec:conclusion} concludes paper and suggest possible directions for future research.
\end{itemize}

\section{Background and related works}
For a given weight matrix $W$ for a graph $G=(V,E)$ with the set of vertices $V$ and edges $E$, the unnormalized Laplacian is defined as $L=D-W$ while the normalized symmetric version is given by $L = I-D^{-1/2}WD^{-1/2}$. Here, $D$ is the diagonal matrix with elements defined as $d_{ii}=\sum^N_{j=1} w_{ij}$ where $w_{ij}$ are the elements of $W$.The Laplacian operator captures many essential graph properties \cite{chung1997spectral,merris1994laplacian}  and find applications in many areas \cite{spielman2010algorithms} such as the spectral clustering \cite{von2007tutorial} or its quantum version \cite{daskin2017quantum}. 

Geometrically motivated data analysis approaches typically relies on the assumption that high-dimensional data lies on a low-dimensional manifold \cite{belkin2006convergence}. In this framework, the graph Laplacian corresponds to the Laplacian differential operator on the manifold \cite{meilua2024manifold}.  While principal component analysis approximate a high dimensional data by identifying a lower dimensional subspace, nonlinear techniques such as the Laplacian eigenmaps transform the data into a lower-dimensional subspace while preserving the geometric relationships. 
Thus, many properties of the graph can be understood and manipulated through the eigenspectrum of the Laplacian \cite{abbas2021power,meilua2024manifold}.
For instance, for the graph Laplacian, eigenvectors form a Fourier-like basis, with eigenvalues corresponding to frequencies. Therefore, a low-pass spectral filter can be used to smooth signals by attenuating high-frequency components \cite{ramakrishna2020user}.

The normalized Laplacian matrix is symmetric; therefore, it can be shown in its eigendecomposition as $L=U\Lambda U^T$, where $\Lambda$ is the eigenvalue matrix with the diagonal elements $\left(\lambda_0, \dots, \lambda_{N-1}\right)$ and $U$ represents the matrix whose columns are the eigenvectors. We will assume that eigenvalues are ordered; therefore,  $\lambda_j \leq \lambda_{j+1}$ and  $\lambda_{N-1} = \lambda_{max}$. Note that the  eigenvalues of a normalized Laplacian matrix are confined to the interval $[0, 2]$ while those of an unnormalized Laplacian matrix lie in $[0, \infty)$. For the remainder of this paper, we will use the normalized Laplacian matrix.

Convolving a function by using another function is generally called convolution. For images, the convolution is performed by using a small matrix, the filter kernel, to generate a filtered image. Using different filters, various operations such as sharpening, blurring can be performed on images.
The convolution neural networks \cite{lecun2002gradient} are designed by integrating convolutions with the Euclidean graph of the image as neural network layers (see Ref.~\cite{o2015introduction,wu2017introduction} for an explanatory introduction on CNNs and the surveys \cite{li2021survey,chen2024review} for their applications).   For images, while the definition of the convolution of a filter kernel is straightforward for the graph representation of the images, for the arbitrary structured graphs, spectral graph convolutions can be defined in the eigendomain of the Laplacian matrices \cite{defferrard2016convolutional,he2022convolutional} (see Ref.~\cite{ghojogh2024graph} for a simple intro). 

The eigenvalues and associated eigenvectors of the Laplacian matrix represent the frequencies and the graph Fourier-basis, respectively. Therefore, the eigendecomposition of the Laplacian provides a way to manipulate the graph in the spectral domain, analogous to the classical Fourier transform \cite{ghojogh2024graph}: For instance $f(\vec{x}) = U^T\vec{x}$ gives the graph Fourier transform for input signal $\vec{x}$.
Frequency domain provides an efficient way to compute the graph convolutions for two signals $\vec{x}$ and $\vec{g}$ because of operator duality between  convolution in the vertex (spatial) domain and the element-wise multiplication, $\odot$, in the spectral domain \cite{rippel2015spectral,defferrard2016convolutional}:
\begin{equation}
    \vec{x}*\vec{g} \iff f(x)\odot f(g).
\end{equation}
Therefore, the convolution in the spectral domain can be written as: 
\begin{equation}
    (U^T\vec{x}) \odot (U^T\vec{g}) = \vec{g}(\Lambda) U^T \vec{x}.
\end{equation}
Applying the inverse transform brings the above convolution result back to the vertex domain:
\begin{equation}
   \vec{x} *\vec{g} = U \vec{g}(\Lambda)U^T x,
\end{equation}
If every node has a feature vector, then the input becomes a matrix $X$ and  can similarly be represented as:
\begin{equation}
    \vec{x}*\vec{g} = U \vec{g}(\Lambda) U^T X.
\end{equation}
In practice, $\vec{g}(\Lambda)$ is not explicitly computed; instead, its directly parameterized as a function of the Laplacian with learnable parameters $\theta$.
Different filters can be defined by using the different functions of $L$. For instance, an order $K$ parameterized polynomial filter $\vec{g}_\theta(L)$ can be defined as \cite{kenlay2020stability}:
\begin{equation}
    \vec{g}_\theta(L)\vec{x} = \sum^K_k \theta_k L^k \vec{x} = \sum^K_k \theta_k U \Lambda^k  U^T \vec{x}.
\end{equation}
As a function of the Laplacian matrix, this operation is equivalent to the functions of its eigenvalues, and hence requires the exact eigendecomposition (or the powers) of the Laplacian matrix. Obtaining the full eigenspectrum of a matrix has a computational cost of $O(N^3)$  for general matrices of dimension $N$ \cite{golub2013matrix}. This is computationally difficult when the given graph is large. It has been shown that this computational cost can be made linear in the number of edges by using polynomial approximation techniques such as Chebyshev polynomials \cite{hammond2011wavelets}:
\begin{equation}
\vec{g}_\theta(L)\vec{x} = \sum_k^KT_k(\tilde{L})\vec{x},   
\end{equation}
where $\tilde{L} = 2L/\lambda_{max}-I$, so that the eigenvalues of $\tilde{L}$ are scaled to the region [-1,1].
CNNs presented in Ref.~\cite{defferrard2016convolutional} employ this form of convolutions by computing $T_k(\tilde{L})$ through recursive computation. There are also other convolutional neural networks that employ different approximations, such as TaylorNet \cite{xu2024taylornet} which uses Taylor expansion and Ref.~\cite{huang2021revisiting} which also employs Laguerre and Hermite polynomials in addition to the Chebyshev polynomials for the Laplace–Beltrami operator of the graph.

\subsection{Related works}
\label{sec:relatedworks}
There are many studies on the implementation of quantum convolutional neural networks and quantum graph neural networks (see Ref.~\cite{ceschini2024graphs} for a critical review and a brief earlier survey \cite{yu2023quantum}).
These models generally translate the core parts of classical CNNs into the quantum domain by replacing convolutional filters and pooling operations with parameterized quantum circuits and/or measurements on the qubits. As in many quantum machine learning and optimization algorithms, the parameters are optimized through a classical optimization algorithm. 
Among these works, QCNN proposed in Ref.~\cite{cong2019quantum} uses variational circuits with $\log(n)$ number of gates for an input of $n$ qubits and shows that the circuit can be understood through multiscale entanglement renormalization ansatz and quantum error correction. While Ref.~\cite{oh2020tutorial} explains and applies the variational circuit types of Ref.~\cite{cong2019quantum} for image classification, Ref.~\cite{chen2022quantum} applies them to high-energy physics problems, and Ref.~\cite{herrmann2022realizing} applies them to recognizing quantum phases. 
Ref.~\cite{umeano2023can} investigates the learnability of various problems using similar types of circuits. Refs.~\cite{liu2021hybrid,smaldone2023quantum,kim2023classical,gong2024quantum} also describe a QCNN model similar to the variational circuit of Ref.~\cite{cong2019quantum} and apply it to different tasks.
Ref.~\cite{kerenidis2019quantum} shows how convolutions can be carried out on quantum computers by uploading data from quantum RAM and using tensor-network representations, presenting a generic framework for QCNN using shallow quantum circuits. Ref.~\cite{parthasarathy2021quantum} shows a quantum version of optical neural networks that uses singular value decomposition.
Ref.~\cite{kerenidis2019quantum} shows how convolutions can be carried out on quantum computers by uploading data from quantum RAM and using tensor-network representations, presenting a generic framework for QCNN using shallow quantum circuits. Ref.~\cite{parthasarathy2021quantum} shows a quantum version of optical neural networks that uses singular value decomposition.
Ref.~\cite{bermejo2024quantum} shows that the described QCNNs are classically simulable, since they are based on shallow circuits for up to 1024 qubits (see Ref.~\cite{napp2022efficient} for efficient classical simulation of the quantum circuits).

Quantum graph neural networks (QGNNs) are also mostly based on various parameterized circuit frameworks. Ref.~\cite{verdon2019quantum} describes an ansatz that uses the time-ordered product of exponentials of Hamiltonians with learnable parameters. It then specifies this ansatz as quantum graph recurrent neural networks by adding temporal parameters between iterations, and as QGNNs by introducing global trainable parameters analogous to the permutation invariance of classical GNNs. Additionally, drawing inspiration from the quantum approximate optimization algorithm \cite{farhi2014quantum} and the alternating operator ansatz \cite{hadfield2019quantum}, it shows that by using alternating coupling and kinetic Hamiltonian operators, Laplacian-based graph convolutional networks can be defined. 
Ref.~\cite{zhang2019quantum} describes quantum walk–based convolutional neural networks that achieve good results on many datasets by capturing both global and local structures of graphs. Ref.~\cite{hu2022design} presents QGCNs with variational circuits based on Givens (plane) rotations. Ref.~\cite{zheng2024quantum} provides quantum circuits for spatial convolution and pooling layers. Similar architectures have been applied to various problems, such as financial fraud detection \cite{innan2024financial} and materials search \cite{ryu2023quantum}. Ref.~\cite{daskin2024unifying} proposes that QGNN models can be unified under the framework of parameterized graph states, which represent graph data as quantum states.
A recent review article, Ref.~\cite{ceschini2024graphs}, provides a systematic review of quantum graph neural networks along with brief discussions of the proposed approaches.

In addition, there are quantum feature‐map extraction schemes \cite{umeano2024ground,matsumoto2025iterative,dou2023efficient} proposed to be used along with CNNs. A more similar work to our approach is Ref.~\cite{chen2021hybrid}, which uses a variational circuit with amplitude embedding and a convolution scheme based on the adjacency matrix to define hybrid quantum CNNs tested on high-energy physics data. The difference in our approach is that we use the parameterized quantum Fourier transform and qubit connections described for a hybrid graph neural network as explained in the Methods.
Note that there is also an earlier work \cite{turek2013quantum} that uses the physical realization of a quantum particle as a graph and implements a spectral filter on it and there are recent studies using quantum Fourier based circuits for the quantum simulation of antisymmetric matrices \cite{daskin2025quantum} and  for finding partial differential equations in Fourier space \cite{lubasch2025quantum}.

\section{Methods}
\label{sec:methods}
To implement a spectral filter on quantum computers, we first show that the eigenspace of the normalized operator \(L\) can be approximated with high accuracy by using a variational quantum circuit whose single layer is defined as follows (also shown in Fig.~\ref{fig:1layercircuit}):
\begin{itemize}
    \item A layer of rotation-\(y\) and controlled-\(y\) gates, which can be considered as permuting and rotating the eigenspace.
    \item A parameterized quantum Fourier transform circuit.
\end{itemize}
For a generic Laplacian operator, the eigendecomposition is as difficult as that of any other symmetric operator. However, Laplacian operators have eigenspaces analogous to the classical Fourier transform. For special graphs with cycles or certain structures, the eigendecomposition of the Laplacian matrices can be defined analytically.
Since the trigonometric nature of Laplacian eigenvectors similar in nature to quantum Fourier bases, the parameterized QFT layers naturally model these oscillatory modes when initialized with spectral priors based on the connections of the graph.

It is well known that the eigenvalues of a Hamiltonian written as a linear combination of tensor products of Pauli matrices can be efficiently estimated using a variational eigensolver \cite{peruzzo2014variational}. The above variational framework in matrix form is very similar to the eigenspace of a Laplacian operator. Therefore, these background and the numerical simulation results shown in Fig.~\ref{fig:pqft-for-different-qubits} indicate that this circuit framework can be used to approximate the eigenspace of any graph Laplacian by using the correct parameters, i.e., the number of layers and connection patterns, which are explained next.

\subsection{Incorporating graph connection patterns into the circuit}

To accommodate connection patterns of the graph in the parameterized circuit, we use the following connection scheme:
For a given graph with \(N\) nodes, using a quantum circuit with \(N\) qubits makes it easy to map the Laplacian operator into the circuit. However, an \(N\)-qubit circuit is not practical with current quantum computer technologies. Therefore, we use \(n\) qubits and describe a contraction method based on the binary representations of the nodes, which generates an \(n \times n\) qubit connection matrix \(M\) with elements defined as:

\begin{align}
\label{eq:connectionM}
M[c,t] &= 
\begin{cases}
   M[c,t] + \frac{A[i,j]}{N}, & \text{if } (i)_2[c] = 1 \text{ and } (j)_2[t] = 1,\\
   0, & \text{otherwise}.
\end{cases}
\end{align}
Here, the entry \(A[i,j]\) represents a connection between the qubits related to indices \(i\) and \(j\). The notation \((\cdot)_2\) denotes the binary string of the given integer. The intuition for \(M\) comes from the symmetry of the adjacency matrix, \(A[i,j] = A[j,i]\), and the fact that we apply a controlled gate when the control qubit (represented by \(c\)) is 1. For asymmetric (directed) graphs, one can similarly check the bit value of the control.
Note that, in addition to the adjacency matrix, the connection matrix can be generated from a weight matrix or via a different mapping method or function. The particular purpose is to map a graph with \(N\) nodes to \(n\) qubits, where \(n \ll N\).

\begin{figure*}
    \centering
    \includegraphics[width=1\linewidth]{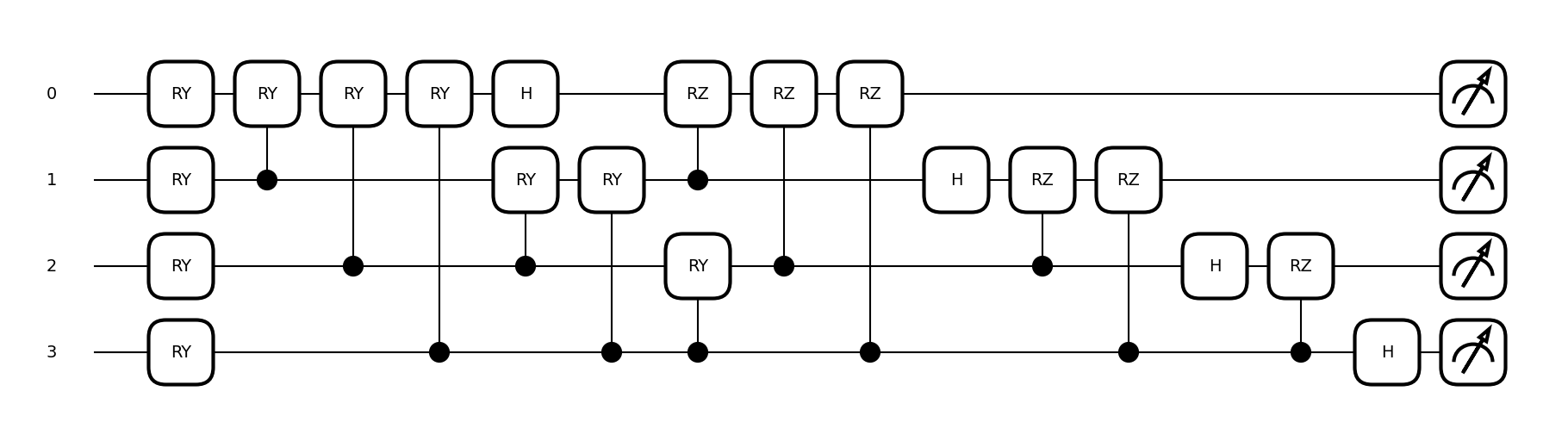}
    \caption{One layer of the four-qubit parameterized connection-based quantum Fourier circuit for approximating eigenspaces of graph Laplacians. When the adjacency matrix entry $A(i,j)\neq 0$, the $control$ and $target$ bit values are checked in the binaries $(i)_2$ and $(j)_2$. If both are '1', then the $(control, target)$ entry of the qubit connection matrix is  updated by the value $A(i,j)$.}
    \label{fig:1layercircuit}
\end{figure*}

Fig.~\ref{fig:1layercircuit} represents a layer of the circuit for a graph with dense qubit connection matrix. For sparse connection matrices, the connecting gates (the entangling gates) $CRY$ and $CRZ$ between qubits  exist if the related entry in the connection matrix is nonzero.

The initial phase parameter values for the $CRZ$ gates inside QFT circuits are also determined using the matrix $M$. For $CRZ(phase_{c,t})$ on qubit$-t$ controlled by the qubit$-c$, the phase determined as:
\begin{equation}
    phase_{c,t} = \left(1-\alpha_{init}\right) rand_{[c,t]} + \alpha_{init} M[c,t], 
\end{equation}
where $\alpha_{init}$ is a ratio parameter and $rand_{[c,t]}$ is a random phase. We use $\alpha_{init}=0.5$ in our optimization  for simplicity. Based on the problem choice, a different value can be chosen or $\alpha_{init}$ can be treated as a learnable parameter. It is important to note that, in some cases when learning multiple $L$ matrices,  $\alpha_{init}$ should be chosen small to avoid overfitting. 
Therefore, in our experiments in Sec.~\ref{sec:experiments}, we use $\alpha_{init}=0.1$ and add a small noise to all matrix elements to prevent overfitting.

Note that when  edge attributes exist, instead of the adjacency matrix one can use a weight matrix $W$ incorporating those attributes. In this paper, we ignore these attributes in our model.
\subsubsection{Limitations of this approach and possible remedies} 
It is possible to face issues when the graph is too sparse, so that the adjacency matrix has a row or column of zeros (isolated nodes). In addition, this structure leads to an uninformative qubit connection matrix: for instance, when there are only 1s in the first row/column of the matrix, the phase connection is 0. This can cause the phase matrix (used for $CRZ$ gates) to be zero, rendering the parameterized QFT ineffective.

For graphs that require a large number of qubits, the optimization performance is limited by the expressivity-power of the circuit. For a generic, non-structured, and full rank matrix $L$ with $k$ independent parameters, a circuit with a number of parameters polynomially close to $k$ is needed to approximate the full eigenspace of the matrix.  
This limitation can be remedied by targeting only a certain part of the eigenspectrum instead of the full spectrum or by simply increasing the number of layers.

\subsection{The circuit expressivity and required number of layers}
\begin{figure*}
\begin{subfigure}[t]{0.5\textwidth}
        \centering
    \includegraphics[width=1\linewidth]{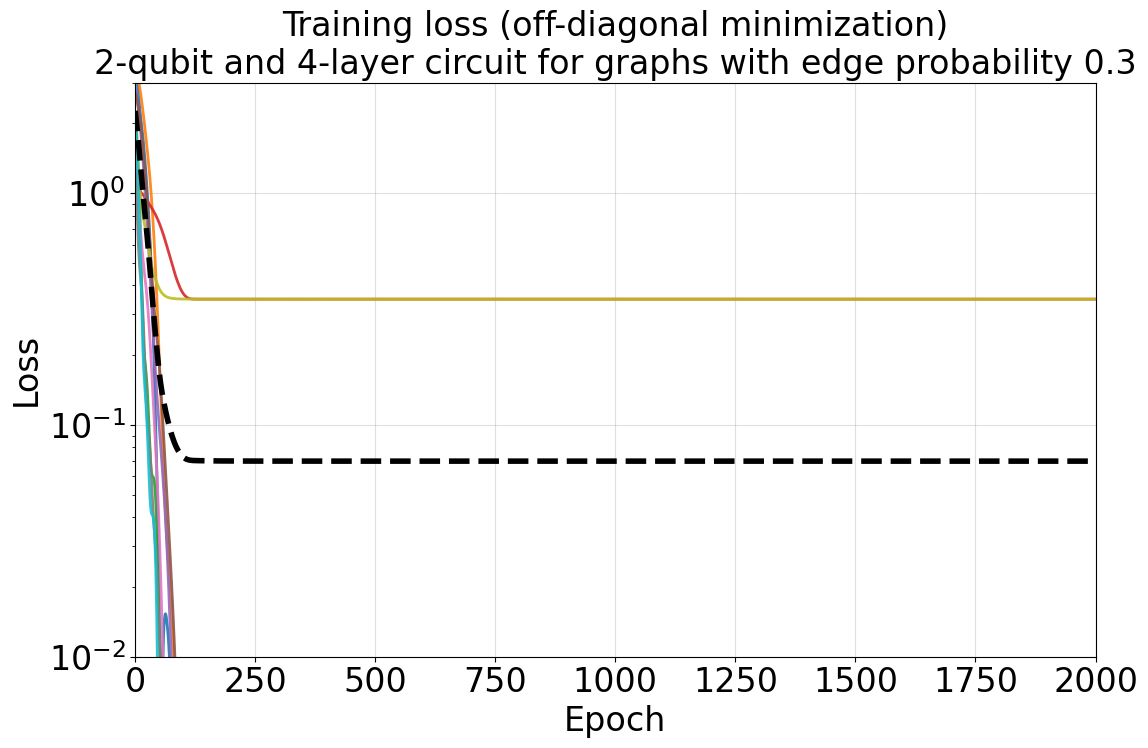}
    \caption{\label{fig:pqft-for-different-qubits2}}
\end{subfigure}~
\begin{subfigure}[t]{0.5\textwidth}
        \centering
    \includegraphics[width=1\linewidth]{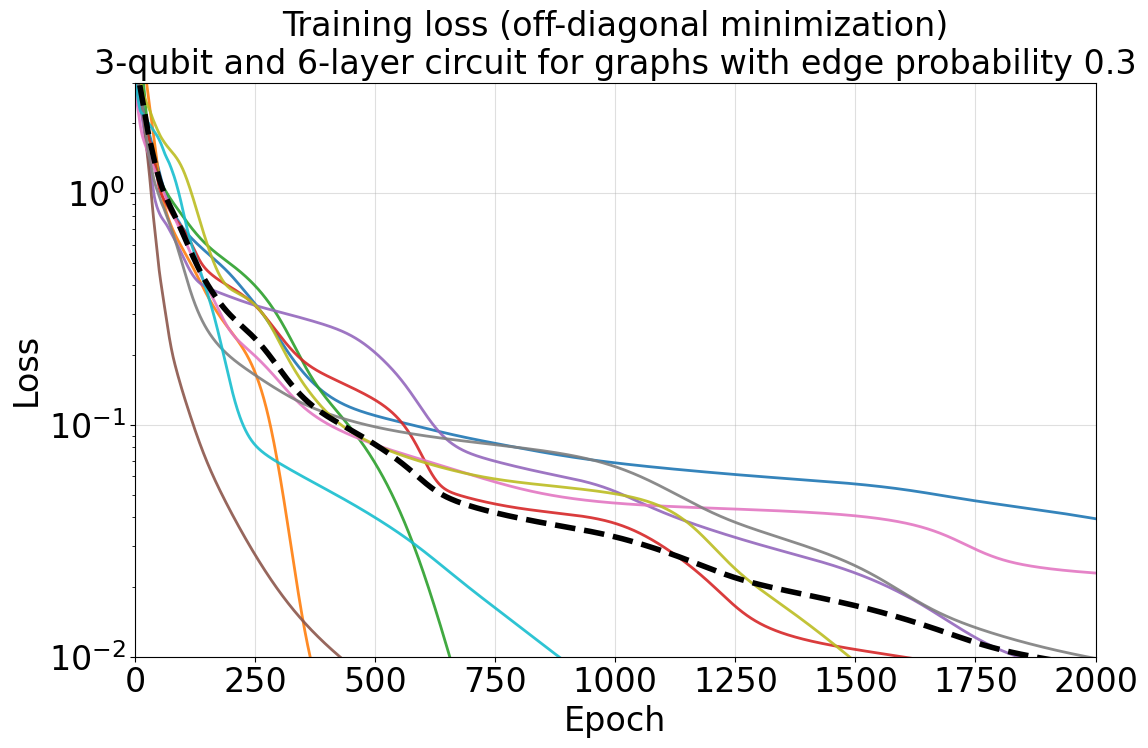}
    \caption{\label{fig:pqft-for-different-qubits3}}
\end{subfigure}\\
\begin{subfigure}[t]{0.5\textwidth}
        \centering
    \includegraphics[width=1\linewidth]{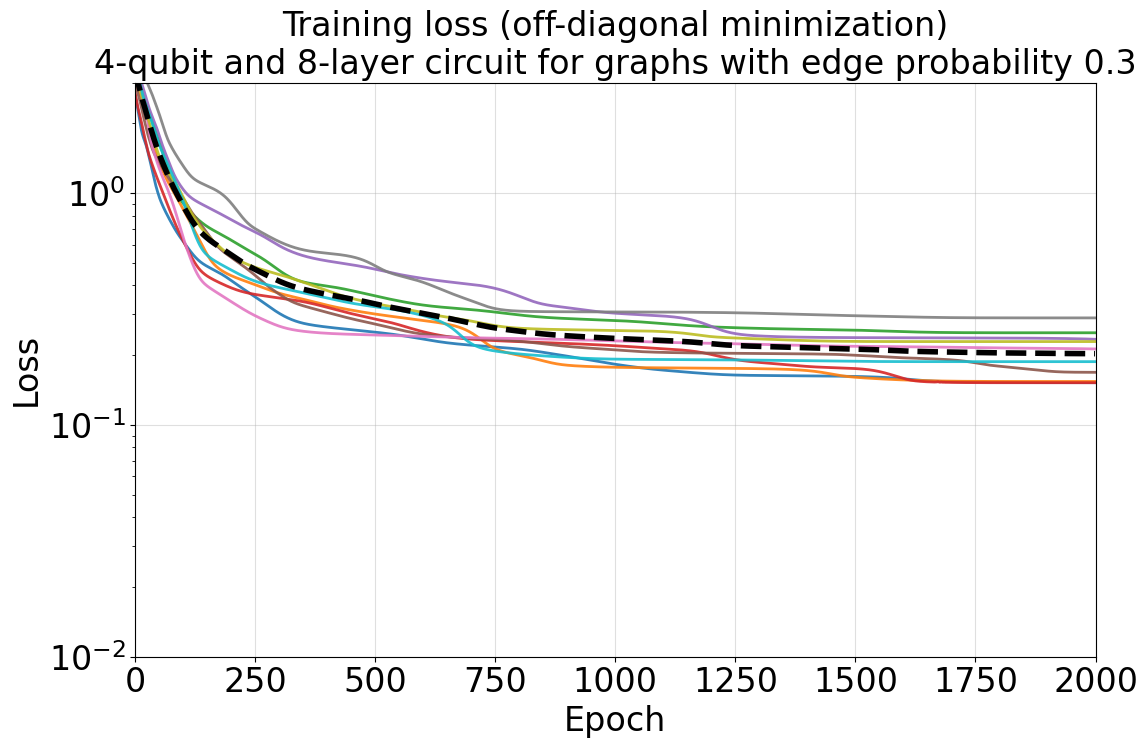}
    \caption{\label{fig:pqft-for-different-qubits4}}
\end{subfigure}~ 
\begin{subfigure}[t]{0.5\textwidth}
        \centering
    \includegraphics[width=1\linewidth]{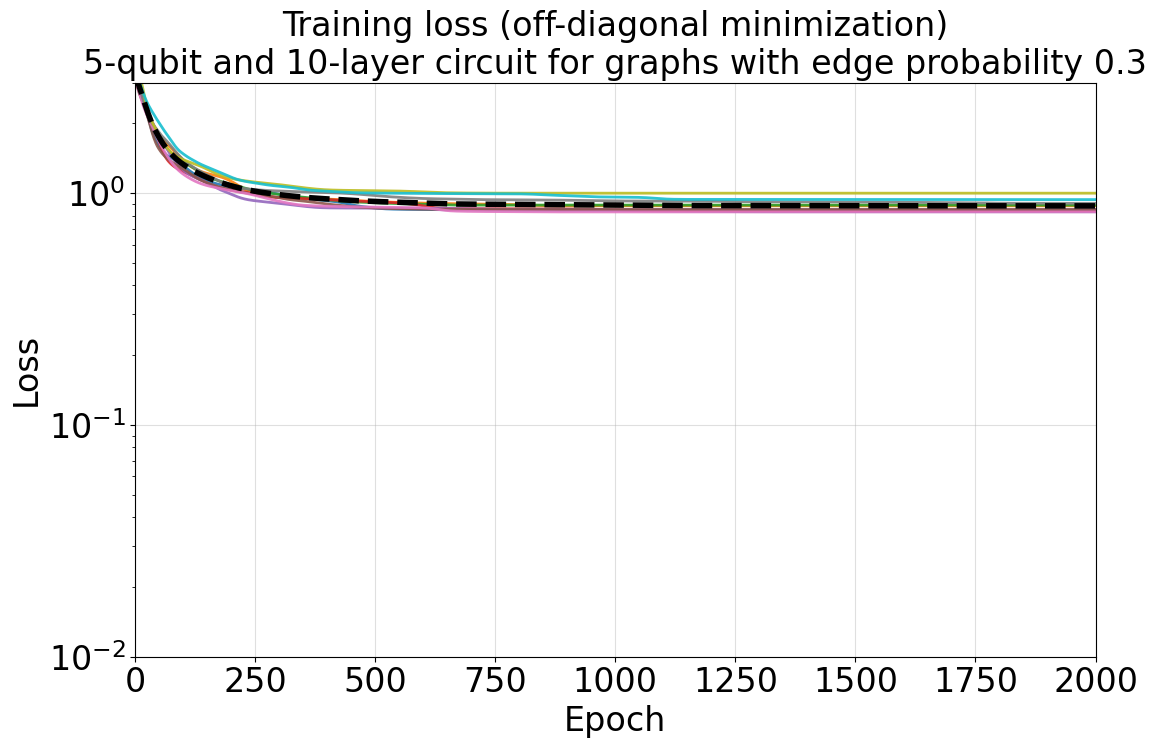}
    \caption{\label{fig:pqft-for-different-qubits5}}
\end{subfigure}\\
\begin{subfigure}[t]{0.5\textwidth}
        \centering
    \includegraphics[width=1\linewidth]{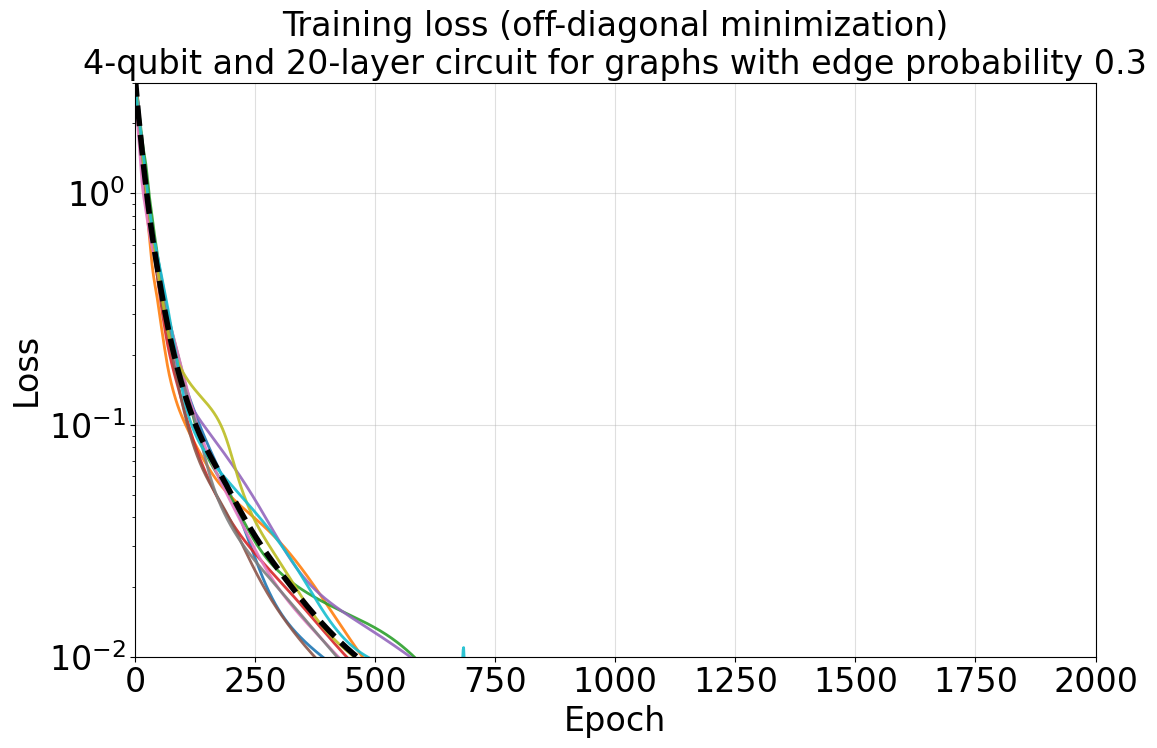}
    \caption{\label{fig:pqft-for-different-qubits4b}}
\end{subfigure} 
\begin{subfigure}[t]{0.5\textwidth}
        \centering
    \includegraphics[width=1\linewidth]{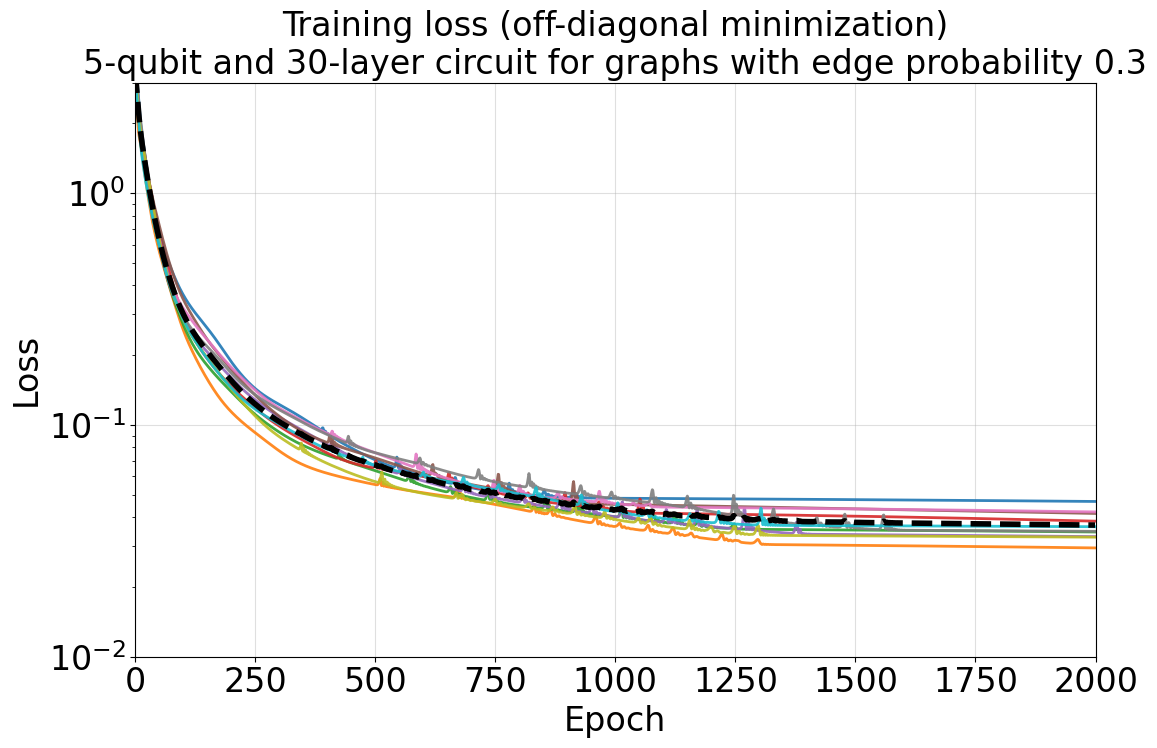}
    \caption{\label{fig:pqft-for-different-qubits5b}}
\end{subfigure}
\caption{The loss function for 10 randomly generated Erdős–Rényi graphs with edge probability 0.3, using various numbers of qubits and layers. The black dashed line shows the mean loss, computed as the squared Frobenius norm of the off‐diagonal elements. As shown in subfigures \ref{fig:pqft-for-different-qubits4b} and \ref{fig:pqft-for-different-qubits5b}, increasing the number of layers enhances the circuit’s expressivity and results in much lower final loss values.}
    \label{fig:pqft-for-different-qubits}
\end{figure*}

To show the approximation power of this framework, in Fig.~\ref{fig:pqft-for-different-qubits} we present simulation results for different numbers of qubits and layers.
During the optimization, the unitary for the circuit \(U_{\text{circuit}}^{(k)}\) is generated for the given parameters at iteration \(k\) by using matrix multiplication:
\begin{equation}
    \hat{\Lambda}^{(k)} = U_{\text{circuit}}^{(k)\,\dagger} \,L\, U_{\text{circuit}}^{(k)}.
\end{equation}

When \(U_{\text{circuit}}\) is the true eigenspace, then \(\hat{\Lambda}\) is a diagonal matrix containing the eigenvalues of \(L\).
The results in the figures show the loss function, calculated using the Frobenius norm of the off-diagonal elements of \(\hat{\Lambda}^{(k)}\):
\begin{equation}
    \mathrm{Loss}_k = \sum_{i=1}^N \sum_{\substack{j=1 \\ j \neq i}}^N \bigl|\hat{\Lambda}_{ij}^{(k)}\bigr|^2.
\end{equation}
The graphs are randomly generated using an edge probability of 0.3. For small graphs, this is likely to create singular adjacency matrices with entire zero rows or columns. Therefore, in these cases the optimization gets stuck, as shown in the top sub-figure for the 2-qubit case.

Looking through the subfigures Fig.~\ref{fig:pqft-for-different-qubits2}–Fig.~\ref{fig:pqft-for-different-qubits4}, one can see that the circuit’s expressive power gradually diminishes. As discussed before, this can be remedied by adding more layers. When the number of layers increases from 8 and 10 to 20, the circuit approximates the eigenspace of the graph Laplacians with very small errors.

It is important to note that, as discussed in the next section, when these circuit frameworks are used as building blocks for learning tasks, adding more layers and increasing the circuit’s expressive power may not be necessary; more importantly, it may cause overfitting and make optimization difficult.

\subsection{Graph neural network}
\begin{figure*}
    \centering
    \includegraphics[width=\linewidth]{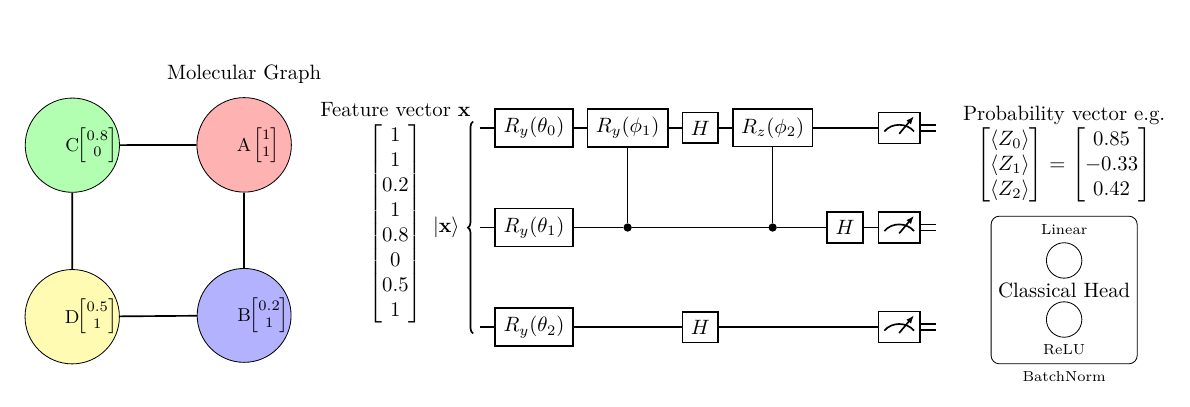}
    \caption{An example workflow of a quantum graph neural network.}
    \label{fig:examplegraph}
\end{figure*}

Here, we describe how the above spectral approximation circuit can be integrated with classical neural networks to design a complete quantum graph neural network framework. Our goal is to show that exponential reduction is achieved while retaining the necessary information to perform various graph tasks. The main framework for our integrated network consists of the following components, which are also visualized in Fig.~\ref{fig:overal-qgnn} and illustrated for a simple graph in Fig.~\ref{fig:examplegraph}:

\begin{itemize}
    \item Graph preprocessing on classical computers. This step generates, for every sample in the dataset:
    \begin{enumerate}[i)]
        \item A feature vector derived from node and edge attributes, which may be provided as matrices or non‐numerical forms. If we have \(d\)-dimensional attributes and \(N\) nodes, this step produces an \(n\log(d)\)-dimensional vector. To match the exact qubit count \(n_q\), we pad the vector with zeros when \(N d\) is not a power of two.
        \item An \(n_q \times n_q\) qubit connection matrix generated using Eq.~\eqref{eq:connectionM}. This matrix indicates which controlled gates are applied for the specific graph instance. When padding the feature vector with zeros, we also pad the adjacency matrix with the corresponding number of zero rows and columns.
    \end{enumerate}
    If graphs vary in size, we apply padding to both adjacency matrices and feature vectors based on the maximum required dimensions.

    \item Circuit structure adaptation. For each sample, we modify the circuit according to the qubit connection matrix: controlled-RY and controlled-RZ gates are applied only if the corresponding matrix entry is nonzero.
    \begin{itemize}
        \item In our experiments, for each phase gate in the quantum Fourier section, we combine 10\% of the calculated phase with a learnable random parameter.
        \item To stabilize training and avoid expressivity loss when using few qubits (we use 4–12 qubits), we add small random noise to the connection matrices to connect every qubit. This prevents overfitting on small datasets.
    \end{itemize}

    \item Quantum pooling. After applying \(l\) layers of the quantum circuit, we measure each qubit in the standard basis to obtain output probabilities, yielding an \(n\)-dimensional vector. Thus, an \(N d\)-dimensional signal for an \(N\)-node graph with \(d\) features per node is reduced to an \(n = \log(Nd)\)-dimensional vector, analogous to pooling in classical CNNs.

    \item Classical prediction head. We then apply a classical neural network composed of layers such as:
    \begin{enumerate}[i)]
        \item A linear transformation: \(\vec{x}' = W^T \vec{x} + b\), where \(W\) is a weight matrix and \(b\) is a bias vector.
        \item A nonlinear activation function, e.g.\ ReLU.
    \end{enumerate}
    This classical multilayer network serves as the prediction head to interpret the circuit’s output probabilities. Its structure can be adapted to the problem’s specific characteristics.
\end{itemize}

Fig.~\ref{fig:examplegraph} shows an example workflow of the above steps.  

\section{Experiments}
\label{sec:experiments}

In our experiments, our main goal is that the quantum–classical hybrid framework can be used in succession for learning purposes, and that the quantum probabilities give enough information to distinguish graphs. For this purpose, we perform graph classification tasks on different datasets listed in the TUDatasets library \cite{morris2020tudataset}.

\textbf{Datasets.} In choosing datasets from the list, we consider the following points:
\begin{itemize}
    \item Since our circuit connections change based on the adjacency matrix, we select datasets where the geometric structure may play an important role.
    \item Since the quantum circuit is simulated, the simulation takes too long for large numbers of samples. Therefore, we choose datasets with a small number of samples so the simulations can run on a desktop computer within a reasonable time.
    \item Since edge attributes require further study (they can be incorporated in a weight matrix or combined within the quantum state vector), we use graph datasets with node attributes and without edge attributes. We leave the study of edge attributes to future work.
\end{itemize}

We select common geometric graph datasets collected from different sources and described for graph learning tasks in Ref.~\cite{Morris+2020,morris2020tudataset} to test model performance on graph classification. The full list of datasets and their properties is given in Table~\ref{tab:datasets}, where we also report the number of qubits based on the feature dimension (including node attributes). The datasets are downloaded using the PyG (PyTorch Geometric) \cite{Fey/Lenssen/2019} library from the TUDatasets.

\textbf{Preprocessing dataset.} The number of qubits is calculated by taking the sample with the highest number of node attributes and features and multiplying by the number of nodes. Then all data is flattened into a vector. Feature vectors and adjacency matrices are padded with the necessary zeros to make the dimensions a power of two and to ensure \(n_q\), the number of qubits, is an integer. Next, we compute the qubit connection matrices and add random noise to each element. No further dataset‐specialized preprocessing is performed.

\textbf{Configurations.} We used PyTorch along with the quantum machine learning package Pennylane for the experiments. The overall network structure and configuration parameters are given in Table~\ref{tab:qgnn_arch}, where \(h_1\) and \(h_2\) are the sizes of the fully connected classical neural network layers with ReLU activation functions. Regularization (BatchNorm) and dropout are used to mitigate possible overfitting.

For optimization, we use the AdamW optimizer with a weight decay of \(1\times10^{-5}\) and reduce the initial learning rate (fixed at 0.01) using PyTorch's \texttt{ReduceLROnPlateau()} scheduler with a factor of 0.1 and patience of 15. Finally, we use the cross-entropy loss function via \texttt{CrossEntropyLoss()}.

\begin{table*}[h]
\centering
\caption{Quantum Graph Neural Network architecture}
\label{tab:qgnn_arch}
\resizebox{\textwidth}{!}
{%
\begin{tabular}{>{\raggedright\arraybackslash}p{0.33\linewidth}>{\raggedright\arraybackslash}p{0.33\linewidth}>{\centering\arraybackslash}p{0.33\linewidth}}
\hline
\textbf{Layer Type} & \textbf{Parameters} & \textbf{Output Dim} \\
\hline
\multicolumn{3}{c}{\textit{Quantum Circuit}} \\
Quantum Spectral Filter & $\#$Qubits: $n_q$, $\#$Layers: $n_\text{layers}$& $n_q$ \\
\hline
\multicolumn{3}{c}{\textit{Classical Head}} \\
Linear & $W \in \mathbb{R}^{n_q \times h_1}$, $b \in \mathbb{R}^{h_1}$& $h_1$ \\
BatchNorm & $h_1$ $\#$features, $\epsilon=10^{-5}$, $\gamma, \beta \in \mathbb{R}^{h_1}$ & $h_1$ \\
ReLU & & $h_1$ \\
Dropout & $p=0.25$ & $h_1$ \\
Linear & $W \in \mathbb{R}^{h_1 \times h_2}$, $b \in \mathbb{R}^{h_2}$ & $h_2$ \\
BatchNorm & $h_2$ $\#$features, $\epsilon=10^{-5}$, $\gamma, \beta \in \mathbb{R}^{h_2}$ & $h_2$ \\
ReLU & & $h_2$ \\
Dropout & $p=0.25$ & $h_2$ \\
Linear & $W \in \mathbb{R}^{h_2 \times n_\text{classes}}$, $b \in \mathbb{R}^{n_\text{classes}}$ & $n_\text{classes}$ \\
\hline
\end{tabular}
}
\end{table*}

\begin{table*}[ht]
\centering
\caption{Required number of qubits for the datasets and their properties taken from TUDatasets \cite{Morris+2020}}.
\label{tab:datasets}
\resizebox{\textwidth}{!}
{%
\begin{tabular}{>{\raggedright\arraybackslash}p{0.15\linewidth}>{\centering\arraybackslash}p{0.15\linewidth}>{\centering\arraybackslash}p{0.15\linewidth}>{\centering\arraybackslash}p{0.1\linewidth}>{\centering\arraybackslash}p{0.15\linewidth}>{\centering\arraybackslash}p{0.15\linewidth}ll}
\toprule
\textbf{Dataset} & \textbf{Type} & \textbf{$\#$Graphs}& \textbf{$\#$Classes}& \textbf{Avg. $\#$Nodes}& \textbf{Avg. $\#$Edges} &\textbf{Max $\#$nodes} &\textbf{$\#$qubits}\\
AIDS & Molecular & 2,000 & 2 & 15.69& 16.20 &95 &12\\
BZR & Molecular & 405 & 2 & 35.75& 38.36 &57 &12\\
COX2 & Molecular & 467 & 2 & 41.22 & 43.45 &56 &12\\
 DHFR& Molecular & 756& 2& 42.43 & 44.54  &71 &12\\
ENZYMES& Molecular & 600& 6 & 32.63 & 62.14 &126 &12\\
MUTAG & Molecular & 188 & 2 & 17.93 & 19.79 &28 &8\\
PROTEINS& Molecular & 1113& 2 & 39.31 & 77.35 &620 &12\\
 PROTEINS\_full& Molecular & 1113& 2 & 39.06 & 72.82 &620 &15\\
COIL-DEL& Geometric& 3900 & 100& 21.54 & 54.24 & 77&8\\
Letter-high & Geometric & 2250 & 15 & 4.67 & 4.50 &9 &5\\
Letter-med & Geometric & 2250 & 15 & 4.67 & 3.21 &9 &5\\
Letter-low & Geometric & 2250 & 15 & 4.68 & 3.13 &8 &4\\
MSRC\_9 & Geometric &    221 & 6& 40.58 & 97.94 &55 &10\\
\bottomrule
\end{tabular}
}
\end{table*}

\begin{table*}[ht]
\centering
\caption{10-fold CV graph classification test accuracy values (\% mean $\pm$ std) on benchmark datasets and comparison to baseline results. GNN-QSF (our model)  parameters: $n_\text{layers}$ quantum  layers, $n_q$ qubits, $[h_1, h_2]$ classical head dimensions, $bs$ is batch size. $\#$params is the number of total parameters: quantum ($\approx n_q^2$) + classical network parameters. The initial learning rate is 0.01 and maximum epoch is 50 for all the runs. Bold indicates better performance.}
\label{tab:results}
\resizebox{\textwidth}{!}{%
\begin{tabular}{lcc>{\raggedright\arraybackslash}p{0.1\linewidth}llcllllclll}
\toprule
\textbf{Dataset} & \textbf{$n_\text{layers}$}& \textbf{$n_q$}&  \textbf{$[h_1, h_2]$}&  bs&\textbf{\#params}&\textbf{GNN-QSF (ours)}&  \multicolumn{5}{c}{\textbf{Baselines from \cite{errica2019fair,pei2024saliency,limbeck2025geometry,you2021identity}}}&&&\\
 & & & &   &&&  GraphSage& GIN& ECC& DiffPool&DGCN& GCN& & \\
AIDS & 4& 12&  [32, 16]&  32&1650&\textbf{0.9965 ± 0.0039}&  & & & 98.8 ± 0.2&&    &&\\
BZR & 4& 12&  [32, 16]&  16&1650&0.8151 ± 0.0578 &  0.852±0.04& 0.856±0.02 & & 76.8 ± 10.1&&    0.844±0.04&&\\
COX2 & 4& 12&  [32, 16]&  32&1650&0.7535 ± 0.0528 &  & & & 76.9 ± 6.5&&    &&\\
MUTAG & 4& 8&  [32, 16]&  16&1202&\textbf{0.8465 ± 0.0790} &  75.8 ± 7.8& 81.4 ± 6.6& 81.4 ± 7.6& 83.3 ± 2.8&83.9 ± 5.8&    71.6 ± 10.9&&\\
MUTAG & \textbf{1}& 8&  [32, 16]&  16&\textbf{1010}&\textbf{0.8412 ± 0.0773} &  75.8 ± 7.8& 81.4 ± 6.6& 81.4 ± 7.6& 83.3 ± 2.8&83.9 ± 5.8&    71.6 ± 10.9&&\\
 PROTEINS& 4& 12& [32, 16]&  32&1650& 0.6703 ± 0.0457& 73.0 ± 4.5& 73.3 ± 4.0& 72.3 ± 3.4& 73.7 ± 3.5& 72.9 ± 3.5& 73.3 ± 3.1& &\\
  PROTEINS\_full& 4& 15& [32, 16]&  32&2070& 0.7125 ± 0.0361& 73.0 ± 4.5& 73.3 ± 4.0& 72.3 ± 3.4& 73.7 ± 3.5& 72.9 ± 3.5& 73.3 ± 3.1& &\\
 ENZYMES& 4& 12& [32, 16]& 32 &1718&  0.3300 ± 0.1035& 58.2 ± 6.0& 59.6 ± 4.5& 29.5 ± 8.2& 59.5 ± 5.6& 38.9 ± 5.7& 68.1±4.0& &\\
Letter-high & 4& 5&  [32, 16]&  32&1171&\textbf{0.9324 ± 0.0213}&  71.0 ± 2.7& 73.3 ± 2.5& 80.8 ± 1.5& &44.7 ± 2.9&    61.1 ± 2.6&&\\
Letter-med & 4& 5&  [32, 16]&  32&1171&\textbf{0.9227 ± 0.0183 }&  & & & &&    &&\\
Letter-low & 4& 4&  [32, 16]&  32&1103&\textbf{0.9471 ± 0.0158} &  & & & &&    &&\\
DHFR & 4& 12&  [32, 16]&  32&1650&0.7395 ± 0.0506 &   & & & 79.8 ± 3.2&&    &&\\
MSRC\_9 & 4& 10&  [32, 16]&  16&1512 &0.7512 ± 0.0709&  & & & &&    &&\\
\bottomrule
\end{tabular}%
}
\end{table*}

\textbf{The training process and reported results.} The model is trained according to the fair comparison guideline given by Errica \emph{et al.} \cite{errica2019fair}, performing 10-fold cross-validation over up to 50 epochs. The best model is selected based on validation performance. Evaluation uses the trained model on the test data, repeated for each fold with a fixed seed of 42. The mean test accuracy is reported.

\textbf{Baseline for comparison.} The key aspect of our model is the exponential reduction of multi-dimensional graph data with node-edge features into a single probability vector while preserving structural information and all attributes for the classical part. Therefore, we compare results only with earlier works such as GraphSage \cite{hamilton2017inductive}, DGCN \cite{zhang2018end}, GCN \cite{kipf2016semi}, GIN \cite{xu2018powerful}, DiffPool \cite{ying2018hierarchical}, and ECC \cite{simonovsky2017dynamic}, to show that the exponential reduction does not significantly affect accuracy. However, not all 50-epoch test accuracies have been reported for our chosen datasets. Thus, we use values from Errica \emph{et al.} \cite{errica2019fair} for ENZYMES and PROTEINS; from Pei \emph{et al.} \cite{pei2024saliency} for some entries of ENZYMES, PROTEINS, Letter-high, and MUTAG; from Limbeck \emph{et al.} \cite{limbeck2025geometry} for some DiffPool entries; and from You \emph{et al.} \cite{you2021identity} for BZR results.

\textbf{Results and discussion.} The mean test results over 10 runs are shown in Table~\ref{tab:results}. For the baseline approaches, the quantum circuit as a spectral method yields mostly comparable results. The boldface entries indicate better performance. From the table it can be seen that since the data mapping is an exponential reduction of the dimensions, when the data size is small (e.g. required numbers of qubits are 4-5 for Letter datasets), the mapping preserves more information. We also used only 1-layer quantum circuit for MUTAG dataset which did not cause a significant performance degrading. This indicates the circuit expressive power and connection based approach can cover most properties of the dataset.
Note that we use generic settings for each dataset; some results may be improved by simple tricks. For instance, the phase‐inclusion parameter in the qubit connection matrix can be adjusted, or a qubit map can be extracted by applying the Laplacian to a few engineered states and observing qubit changes to determine connectivity. Such designs can leverage the entangling and disentangling power of unitaries \cite{linden2009entangling}, as shown in Ref.~\cite{daskin2025quantum}.

\section{Conclusion and future directions}
\label{sec:conclusion}

In this paper, we have described a learnable quantum spectral filter based on a parameterized quantum Fourier transform circuit. Numerical simulations have shown that by layering this circuit, the eigenspace of a graph Laplacian can be approximated.
When we have a dataset of many graphs, applying this approximation can be considered as extracting commonalities of the graphs from the dataset. Using this intuition, we have shown that this circuit can be combined as a learnable spectral filter with classical neural networks to form powerful and efficient graph neural networks, thanks to the exponential dimension reduction achieved by the quantum circuit. The experiments on many structured datasets show that this reduction does not cause significant information loss.

For future research directions, there are many parts of the described framework that can be studied. One important aspect is that the definition of the qubit connection matrix can be specialized to the dataset rather than used as a generic setting, as done in this paper.
Another issue is the quantum state vector. For simplicity, we put all features and attributes related to edges or nodes into a single state vector. Instead, one can incorporate this more carefully into the circuit. For instance, one can use different registers for each type of feature (which would require more qubits) or apply the circuit separately to each register and combine their results in the classical head.
There are also some training issues that can be studied more thoroughly. One issue is using the same learning rate for the classical and quantum parts, which may cause a learning bottleneck when their optimization dynamics differ.
Finally, note that changing the qubit connections—and thus the circuit architecture—may cause simulations to run more slowly on certain quantum hardware if they require specialized settings.

\section{Data availability}
The simulation code and results used for this paper are publicly available
at: \url{https://github.com/adaskin/gnn-qsf}
\section{Funding}
This project is not funded by any funding agency.

\section{Conflict of interest}
The authors declare no conflict of interest.

\bibliographystyle{unsrt}
\bibliography{main}
\end{document}